\documentclass[aps,pre,reprint,longbibliography]{revtex4-1}

\usepackage{amsmath,amsfonts,bm,hyperref,graphicx,verbatim,mathrsfs,units,enumerate}

\usepackage{color,colordvi}
\definecolor{blue}{rgb}{0,0,1}
\definecolor{grey}{rgb}{0.6,0.6,0.6}

\begin{document}

\title{Thermodynamic uncertainty relations including measurement and feedback}

\author{Patrick P. Potts}
\email{patrick.potts@teorfys.lu.se}
\author{Peter Samuelsson}
\affiliation{Physics Department and NanoLund, Lund University, Box 118,  22100 Lund, Sweden.}
%\thanks{}

\date{\today}

\begin{abstract}
Thermodynamic uncertainty relations quantify how the signal-to-noise ratio of a given observable is constrained by dissipation. Fluctuation relations generalize the second law of thermodynamics to stochastic processes. We show that any fluctuation relation directly implies a thermodynamic uncertainty relation, considerably increasing their range of applicability. In particular, we extend thermodynamic uncertainty relations to scenarios which include measurement and feedback. Since feedback generally breaks time-reversal invariance, the uncertainty relations involve quantities averaged over the forward and the backward experiment defined by the associated fluctuation relation. This implies that the signal-to-noise ratio of a given experiment can in principle become arbitrarily large as long as the corresponding backward experiment compensates, e.g. by being sufficiently noisy. We illustrate our results with the Szilard engine as well as work extraction by free energy reduction in a quantum dot.
\end{abstract}

%\keywords{}

\maketitle

%%%%%%%%%%%%%%%%%%%%%%%%%%%%%%%%%%%%%%%%%%%%%%%%%%%%%%
%%%%%%%%%%%%%%%%%%%%%%%%%%%%%%%%%%%%%%%%%%%%%%%%%%%%%%


\section{Introduction} The field of stochastic thermodynamics investigates small, fluctuating systems that are out of equilibrium \cite{seifert:2012,ciliberto:2017,seifert:2018}, with applications ranging from biological \cite{ritort:2008,barato:2016,gnesotto:2018} and chemical systems \cite{rao:2016} over information processing \cite{sagawa:2014} to nanoelectronic devices \cite{sothmann:2015}. In recent years, powerful relations have been discovered that determine the behavior of small systems far from equilibrium. These include thermodynamic uncertainty relations (TURs) \cite{barato:2015,gingrich:2016} as well as fluctuation relations \cite{jarzynski:2000,esposito:2009rmp,seifert:2012}, both of which have significantly contributed to our understanding of non-equilibrium phenomena (e.g., work extraction using measurement and feedback \cite{sagawa:2012} and biological clocks \cite{marsland:2019}). A particularly interesting application of TURs is the inference of dissipation from the measurement of fluctuating currents \cite{li:2019}. While these relations have mostly been treated independently, a connection between TURs and fluctuation relations was recently established under rather restrictive assumptions \cite{hasegawa:2019,vu:2019,timpanaro:2019}. Here we generalize this connection and show that any fluctuation relation implies the existence of a TUR. This allows for the full wealth of results on fluctuation relations to spread over to TURs, significantly extending their range of applicability.

TURs constrain the signal-to-noise ratio of an observable $\phi$ by the associated entropy production \cite{barato:2015,gingrich:2016,pietzonka:2017,horowitz:2017,pietzonka:2018,seifert:2018}
\begin{equation}
\label{eq:tur1}
\frac{\langle\langle \phi^2\rangle\rangle}{\langle \phi\rangle^2}\geq\frac{2}{\langle \sigma \rangle},
\end{equation}
where $\langle \cdot\rangle$ denotes the ensemble average, $\langle \langle \phi^2\rangle\rangle = \langle \phi^2\rangle-\langle \phi\rangle^2$ the variance, and $\sigma$ denotes the entropy production. This inequality was rigorously proven for current observables in time-homogeneous Markov jump processes with local detailed balance \cite{gingrich:2016,horowitz:2017}. Various extensions of the TUR exist including periodically driven systems \cite{koyuk:2018,barato:2018,barato:2019}, systems with broken time-reversal symmetry in linear response \cite{Macieszczak:2018} and beyond \cite{brandner:2018,proesmans:2019}, a tighter version \cite{polettini:2016}, as well as a four times looser bound which holds in the quantum regime for general two-terminal systems \cite{guarnieri:2019}. Of particular interest for our investigation is an extension that was originally derived for discrete-time Markov chains and systems subject to a time-symmetric external driving \cite{proesmans:2017}
\begin{equation}
\label{eq:tur2}
\frac{\langle\langle \phi^2\rangle\rangle}{\langle \phi\rangle^2}\geq\frac{2}{e^{\langle \sigma\rangle}-1},
\end{equation}
where $\sigma$ denotes the entropy produced during one period of the driving.
Note that this bound is less strict than the bound in Eq.~\eqref{eq:tur1}.

Fluctuation theorems are powerful and exact statements which relate probabilities in a forward experiment to probabilities in a backward experiment \cite{jarzynski:2000,seifert:2005,esposito:2009rmp,jarzynski:2011,seifert:2012,mansour:2017,campisi:2011,bochkov:2013}
\begin{equation}
\label{eq:flucrel}
\frac{P_{\rm B}(-\phi,-\sigma_{\rm I})}{P(\phi,\sigma_{\rm I})}=e^{-\sigma_{\rm I}},
\end{equation}
where the subscript B denotes the backward experiment. In its most common version, the backward experiment is obtained by time-reversing the forward experiment, $\phi$ is an observable that is odd under time-reversal, and $\sigma_{\rm I}$ denotes the entropy production \cite{seifert:2012}. There are numerous extensions of the fluctuation relation, illustrating the fact that different choices for the backward experiment result in different extensions of $\sigma_{\rm I}$, each with its own interpretation and merits \cite{esposito:2010three,garcia:2012,seifert:2012,potts:2018,rao:2018}. Here we focus on extensions including measurement and feedback, where $\sigma_{\rm I}$ includes an information term symbolized by the subscript I \cite{sagawa:2008,cao:2009,sagawa:2010,horowitz:2010,ponmurugan:2010,morikuni:2011,sagawa:2012,sagawa:2012prl,sagawa:2013,lahiri:2012,abreu:2012,funo:2013,ashida:2014,horowitz:2014,horowitz:2014prx,funo:2015,wachtler:2016,gong:2016,spinney:2016,spinney:2018,kwon:2017,potts:2018,crooks:2019}.

Very recently, a deep connection between fluctuation relations and TURs has been uncovered. In the special case where Eq.~\eqref{eq:flucrel} holds with $P_{\rm B}=P$, the generalized TUR in Eq.~\eqref{eq:tur2} was shown to hold \cite{hasegawa:2019,vu:2019,turviol}. While this connection extends the regime of applicability of TURs, the constraint $P_{\rm B}=P$ is a strong restriction, preventing the application of TURs in the presence of measurement and feedback, which explicitly breaks time-reversal symmetry. 

Here we lift this constraint and derive a generalized TUR solely from Eq.~\eqref{eq:flucrel}. For $P_{\rm B}\neq P$, we find that the TUR does not directly bound the signal-to-noise ratio in the forward experiment but rather constrains the combination of forward and backward experiment. As a consequence, the signal-to-noise ratio in a given experiment can become arbitrarily large as long as this is compensated by a corresponding backward experiment, as discussed in more detail below. In the absence of measurement and feedback, we recover the results from Ref.~\cite{proesmans:2019}.

\section{The uncertainty relation} We now derive our main result, a generalized TUR, from Eq.~\eqref{eq:flucrel}. Following Refs.~\cite{merhav:2010,hasegawa:2019,vu:2019}, we introduce the auxiliary probability distribution
\begin{equation}
\label{eq:qprob}
\begin{aligned}
Q(\phi,\sigma_{\rm I}) &= \frac{1+e^{-\sigma_{\rm I}}}{2}P(\phi,\sigma_{\rm I})\\&=\frac{1}{2}\left[P(\phi,\sigma_{\rm I})+P_{\rm B}(-\phi,-\sigma_{\rm I})\right],
\end{aligned}
\end{equation}
which is the average of the forward distribution and the backward distribution with negated arguments. Note that in contrast to Refs.~\cite{merhav:2010,hasegawa:2019,vu:2019}, $Q$ is not restricted to positive $\sigma_{\rm I}$. An important property of $Q$ that will be used repeatedly is given by
\begin{equation}
\tanh\left(\frac{\sigma_{\rm I}}{2}\right)Q(\phi,\sigma_{\rm I})=\frac{1}{2}\left[P(\phi,\sigma_{\rm I})-P_{\rm B}(-\phi,-\sigma_{\rm I})\right].
\end{equation}
%This distribution results in the following averages
%\begin{equation}
%\label{eq:avq}
%\left\langle f_{\pm}(\phi,\sigma_{\rm I}) \tanh^{\frac{1\mp1}{2}}\left(\frac{\sigma_{\rm I}}{2}\right)\right\rangle_Q=\overline{\langle f_{\pm}(\phi,\sigma_{\rm I})\rangle},
%\end{equation}
%where
%\begin{equation}
%\label{eq:barparity}
%f_\pm(\phi,\sigma_{\rm I})=\pm f_\pm(-\phi,-\sigma_{\rm I}),\hspace{.5cm}\overline{\langle \cdot\rangle} = \frac{1}{2}\left(\langle \cdot\rangle+\langle \cdot\rangle_{\rm B}\right),
%\end{equation}
We now prove the series of inequalities
\begin{equation}
\label{eq:ineqsder}
\begin{aligned}
\left(\frac{\langle \phi\rangle+\langle\phi\rangle_{\rm B}}{2}\right)^2&=\left\langle\left(\phi-\langle\phi\rangle_Q\right)\tanh\left(\frac{\sigma_{\rm I}}{2}\right)\right\rangle_Q^2
\\&\leq \langle\langle \phi^2\rangle\rangle_Q\left\langle \tanh^2\left(\frac{\sigma_{\rm I}}{2}\right)\right\rangle_Q\\&\leq\langle\langle \phi^2\rangle\rangle_Q\left\langle \tanh\left[\frac{\sigma_{\rm I}}{2}\tanh\left(\frac{\sigma_{\rm I}}{2}\right)\right]\right\rangle_Q\\&\leq\langle\langle \phi^2\rangle\rangle_Q\tanh\left(\frac{\langle\sigma_{\rm I}\rangle+\langle\sigma_{\rm I}\rangle_{\rm B}}{4}\right).
\end{aligned}
\end{equation}
where we introduced the average over $Q$ as $\langle\cdot\rangle_Q$, the average over $P$ as $\langle\cdot\rangle$ and the average over $P_{\rm B}$ as $\langle\cdot\rangle_{\rm B}$.
The first inequality is the Cauchy-Schwarz inequality. The second and third inequalities both use the fact that $\tanh(x)$ is a concave function for $x\geq 0$. This implies the inequality 
\begin{equation}
\label{eq:ineqtanh}
k\tanh(x)\leq\tanh(kx),
\end{equation}
 for $x\geq0$ and $k\in [0,1]$. The second inequality follows by setting $x=|\sigma_{\rm I}|/2$ and $k = \tanh(|\sigma_{\rm I}|/2)$. The third inequality is Jensen's inequality where we made use of 
 \begin{equation}
\langle\sigma_{\rm I}\rangle+\langle\sigma_{\rm I}\rangle_{\rm B}=2\langle \sigma_{\rm I}\tanh(\sigma_{\rm I}/2)\rangle_Q.
 \end{equation}
 By further using
\begin{equation}
\label{eq:varphiq}
\langle\langle \phi^2\rangle\rangle_Q =\frac{1}{2}\left(\langle\langle \phi^2\rangle\rangle+\langle\langle \phi^2\rangle\rangle_{\rm B}\right)+\left(\frac{\langle \phi\rangle+\langle\phi\rangle_{\rm B}}{2}\right)^2,
\end{equation}
we finally obtain
\begin{equation}
\label{eq:gturfb}
\frac{\langle\langle \phi^2\rangle\rangle+\langle\langle \phi^2\rangle\rangle_{\rm B}}{\left(\langle \phi\rangle+\langle \phi\rangle_{\rm B}\right)^2}\geq\frac{1}{\exp\left({\frac{\langle\sigma_{\rm I}\rangle+\langle\sigma_{\rm I}\rangle_{\rm B}}{2}}\right)-1}.
\end{equation}
Equation \eqref{eq:gturfb} is the main result of this paper. It follows directly from Eq.~\eqref{eq:flucrel} and shows how the quantity $\langle\langle \phi^2\rangle\rangle/\langle \phi\rangle^2$ is no longer directly bounded if the backward probability distribution differs from the forward one which is generally the case for broken time-reversal symmetry. In particular, for sufficiently large $\langle\langle \phi^2\rangle\rangle_{\rm B}$ or sufficiently small $\langle \phi\rangle+\langle \phi\rangle_{\rm B}$, the inequality may be respected for arbitrary values for the signal-to-noise ratio in the forward experiment. One can thus try to overcome the traditional uncertainty relation in Eq.~\eqref{eq:tur1} by looking for situations where the backward experiment is noisier than the forward one, or where the signs of the average values differ, see our examples below as well as Ref.~\cite{chun:2019}. 

The TUR in Eq.~\eqref{eq:gturfb} is of the same form as the one found in Ref.~\cite{proesmans:2019} for Markovian systems with broken time-reversal symmetry. However, we stress that as our derivation only requires Eq.~\eqref{eq:flucrel} to hold, Eq.~\eqref{eq:gturfb} is valid for any scenario where a fluctuation relation applies. This includes scenarios with measurement and feedback which has a twofold effect: First, time-reversal symmetry is broken, resulting in the general structure of Eq.~\eqref{eq:gturfb} that includes expectation values of the backward probability distribution. Second, the entropy production $\langle\sigma\rangle$ is modified by an information term $I$ such that $\sigma_{\rm I}=\sigma+I$.

\begin{figure*}
	\centering
	\includegraphics[width=\textwidth]{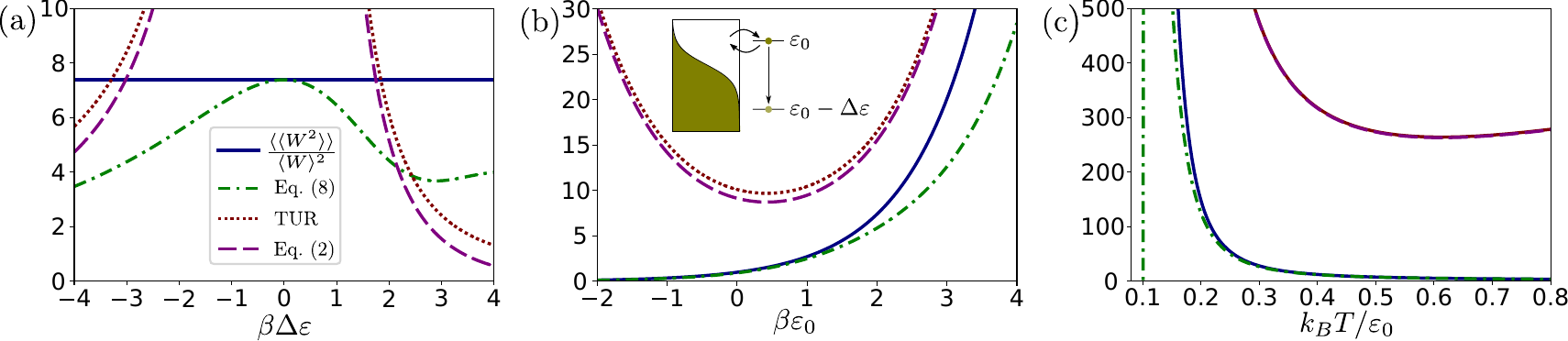}
	\caption{Inverse signal-to-noise ratio for work extraction by instantly lowering the energy level of a quantum dot. The solid, blue line gives the variance of the work divided by the mean squared. The red, dotted line provides the bound by the original TUR [cf.~Eq.~\eqref{eq:tur1}], the purple, dashed line the bound provided by the generalized TUR given in Eq.~\eqref{eq:tur2}. Both of these bounds can be violated since they do not apply to situations with broken time-reversal symmetry. The green, dash-dotted line provides our bound [cf.~Eq.~\eqref{eq:gturstn}] which applies to the present scenario. Inequalities as a function of (a) $\Delta\varepsilon$ ($\beta\varepsilon_0=2$), (b)  $\varepsilon_0$ ($\beta\Delta\varepsilon=1.3$) and (c) temperature ($\Delta\varepsilon/\varepsilon_0=0.16$). Our inequality reduces to an equality in the high temperature limit as well as for $\Delta\varepsilon\rightarrow 0$. We note that in this scenario, our bound is equivalent to the one derived in Ref.~\cite{proesmans:2019}.}
	\label{fig:qdot}
\end{figure*}

\section{Examples}
We will now illustrate Eq.~\eqref{eq:gturfb} with the help of two examples. Details of the calculations are deferred to Apps.~\ref{app:qdot} and \ref{app:szilard}. For illustration purposes, we rewrite Eq.~\eqref{eq:gturfb} such that it represents a lower bound on the signal-to-noise ratio in the forward experiment
\begin{equation}
\label{eq:gturstn}
\frac{\langle\langle \phi^2\rangle\rangle}{\langle \phi\rangle^2}\geq\frac{\left(1+\frac{\langle \phi\rangle_{\rm B}}{\langle \phi\rangle}\right)^2}{\exp\left({\frac{\langle\sigma_{\rm I}\rangle+\langle\sigma_{\rm I}\rangle_{\rm B}}{2}}\right)-1}-\frac{\langle\langle \phi^2\rangle\rangle_{\rm B}}{\langle \phi\rangle^2}.
\end{equation}
We note that the bound provided by the right-hand side depends on the average $\langle\phi\rangle$ but not on the variance $\langle\langle \phi^2\rangle\rangle$.
Before discussing our main example, the Szilard engine, it is illustrative to consider an example without measurement and feedback. To this end, we consider the process of work extraction by lowering the free energy. 

\subsection{Quantum dot}
For concreteness, we consider a quantum dot with a single energy level that is coupled to a fermionic reservoir, see inset in Fig.~\ref{fig:qdot}~(b). The dot is initially in thermal equilibrium with a level energy $\varepsilon_0$ such that its occupation is given by the Fermi-Dirac distribution
\begin{equation}
\label{eq:fermi}
f(\varepsilon_0) = \frac{1}{e^{\beta\varepsilon_0}+1},
\end{equation}
where we set the chemical potential of the reservoir to zero without loss of generality. The energy level is then lowered by the amount $\Delta\varepsilon$. Thereby, the free energy of the system is reduced and energy is extracted. In such a process, a Crooks fluctuation relation \cite{crooks:1999} holds
\begin{equation}
\label{eq:crooks}
\frac{P_{\rm B}(-W)}{P(W)} = e^{-\beta (\Delta F-W)},
\end{equation}
where $\beta=1/(k_BT)$ denotes the inverse temperature. The backward experiment corresponds to initiating the dot in thermal equilibrium at energy $\varepsilon_0-\Delta \varepsilon$ and lifting the energy level to $\varepsilon_0$ in a manner that corresponds to the time-reversal of the forward experiment. In Eq.~\eqref{eq:crooks}, $\Delta F$ is the free energy difference between the initial states of the forward and backward experiments. Equation \eqref{eq:crooks} is of the form of Eq.~\eqref{eq:flucrel}, with $\phi=W$ and $\sigma_{\rm I}=\beta(\Delta F-W)$. We note that a cyclic version of this process was investigated experimentally in Refs.~\cite{hofmann:2016,hofmann:2017}. Here we consider two limiting cases: the quasistatic limit, where the dot is moved infinitely slowly, and the instantaneous limit, where the dot is moved infinitely fast.

In the quasistatic limit, an equivalence between ensemble average and time average implies that the dot can be described as remaining in thermal equilibrium even for a single realization of the experiment, see App.~\ref{app:qdot}. It can then be shown that the amount of work that is extracted equals the reduction in the free energy such that
\begin{equation}
\label{eq:qdotstat}
P(W)=\delta(W-\Delta F),\hspace{.5cm}P_{\rm B}(W)=\delta(\Delta F-W),
\end{equation}
Fulfilling Eq.~\eqref{eq:crooks}. For the averages and variances we thus find
\begin{equation}
\label{eq:avvar}
\langle W\rangle=-\langle W\rangle _{\rm B}=\Delta F,\hspace{.5cm}\langle\langle W^2\rangle\rangle = \langle\langle W^2\rangle\rangle_{\rm B}=0.
\end{equation}
The extracted work in the forward experiment is thus finite but both its variance, as well as the associated entropy production vanish. This is in clear violation of the TURs given in Eqs.~\eqref{eq:tur1} and \eqref{eq:tur2}. However, no contradiction arises with our inequality. Indeed, in this case the right-hand side of Eq.~\eqref{eq:gturstn} reduces to zero, resulting in a trivial inequality because the left-hand side is a positive quantity by definition.
While the quasistatic limit is clearly an idealized situation, the general strategy of reducing $\langle W\rangle+\langle W\rangle _{\rm B}$ in order to obtain large power outputs with small fluctuations and entropy productions provides a promising avenue to pursue.

In the instantaneous limit, the work $\Delta \varepsilon$ is extracted if the dot is initially filled. This happens with probability $f(\varepsilon_0)$. Otherwise, no work is extracted.
This is described by the distributions
\begin{equation}
\label{eq:workdistinst}
\begin{aligned}
P(W) = &\delta(W-\Delta \varepsilon)f_0+\delta(W)[1-f_0],\\
P_{\rm B}(W) =& \delta(W+\Delta \varepsilon)f_1+\delta(W)[1-f_1],
\end{aligned}
\end{equation}
where we abbreviated $f_0=f(\varepsilon_0)$ and $f_1=f(\varepsilon_0-\Delta\varepsilon)$.

In contrast to the quasistatic limit, both the variance of the extracted work, as well as the entropy production are in general finite and we find
\begin{equation}
\begin{aligned}
\label{eq:avgfisnt}
\langle W\rangle &= \Delta\varepsilon f_0,\hspace{.65cm}\langle\langle W^2\rangle\rangle = (\Delta\varepsilon)^2f_0[1-f_0],\\
\langle W\rangle_{\rm B} &= -\Delta\varepsilon f_1,\hspace{.35cm}\langle\langle W^2\rangle\rangle_{\rm B} = (\Delta\varepsilon)^2f_1[1-f_1].
\end{aligned}
\end{equation}
In particular, these expressions result in
\begin{equation}
\label{eq:varwd}
\frac{\langle\langle W^2\rangle\rangle}{\langle W\rangle^2} = e^{\beta\varepsilon_0},
\end{equation}
which is only bounded from below from zero. This expression is shown in Fig.~\ref{fig:qdot}, together with bounds provided by Eqs.~\eqref{eq:tur1}, \eqref{eq:tur2}, and \eqref{eq:gturstn}. We find that both previous bounds can be violated while our bound is always satisfied. We note that in the limit $\beta\varepsilon_0\rightarrow-\infty$, the dot is always occupied and thus remains in thermal equilibrium. We then recover the results for the quasistatic limit. In the limit $\beta\Delta\varepsilon\rightarrow 0$, which includes the high temperature limit as well as the limit where the level is not moved at all, the right hand side of Eq.~\eqref{eq:gturstn} reduces to Eq.~\eqref{eq:varwd} and our bound becomes tight.

\subsection{Szilard engine}
\begin{figure*}
	\centering
	\includegraphics[width=\textwidth]{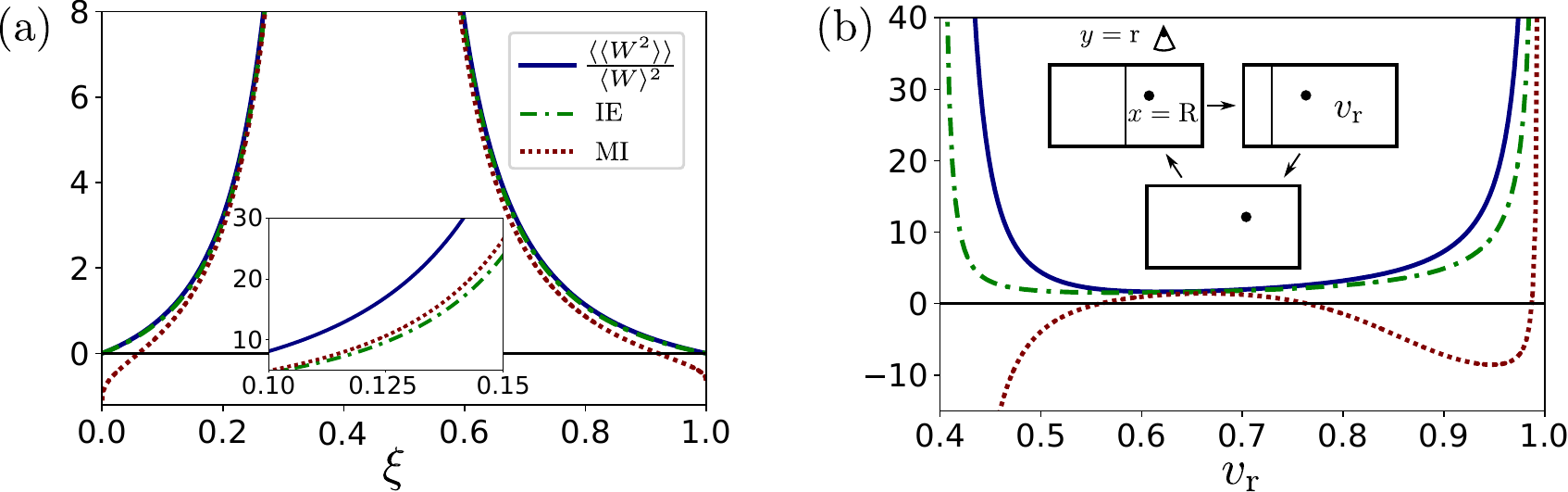}
	\caption{Thermodynamic uncertainty relations for the Szilard engine. The solid, blue line gives the variance of the work divided by the mean squared, the red, dotted line provides the bound that derives from the fluctuation relation where the information term is related to the mutual information \cite{sagawa:2012}, the green, dash-dotted line derives from the fluctuation relation related to the inferable entropy production \cite{potts:2018}. (a) Inequalities as a function of the measurement error for $v_{\rm l}=v_{\rm r}=0.65$. We note that there is a small but finite difference between the blue and the green lines for $\xi\neq0,1$. The diverging feature occurs at the measurement error where the work changes sign. Depending on the parameters, either backward experiment can result in a tighter bound, see inset, where $v_{\rm l}=v_{\rm r}=0.95$. (b) As a function of $v_{\rm r}$, the two bounds show a qualitatively different behavior. Here $v_{\rm l}=0.65$ and $\xi = 0.15$.}
	\label{fig:szilard}
\end{figure*}
Our second example includes measurement and feedback and is provided by the Szilard engine \cite{szilard:1929}. The experiment is sketched in the inset of Fig.~\ref{fig:szilard}\,(b) and includes the following steps
\begin{enumerate}
	\item Prepare a single particle in a box, the volume of which is set to one.
	\item Insert a partition, dividing the box into two parts of equal volume. The variable $x={\rm L,R}$ (left, right) encodes the location of the particle.
	\item Determine the location of the particle. The variable $y={\rm l,r}$ (left, right) encodes the outcome of this measurement. An error of the measurement (e.g., $x={\rm L}$ and $y={\rm r}$) occurs with probability $\xi$.
	\item Slowly move the partition until the part of the box where the particle was measured to be reaches the volume $v_y\leq 1$.
\end{enumerate}
 This procedure is described by the joint probability distribution $P(x,y)$.
 For $v_y>1/2$, and no measurement error, work is extracted from the particle akin to the work that is extracted from an expanding gas, see App.~\ref{app:szilard}. If a measurement error occurs, work is performed on the particle akin to the compression of a gas. The extracted work for given $x$ and $y$ will be denoted $W(x,y)$.

 Here we consider two backward experiments. The first was introduced by Sagawa and Ueda \cite{sagawa:2012} and includes the following steps
 \begin{enumerate}
 	\item Choose a value for $y$ from the marginal probability distribution $P(y)=\sum_xP(x,y)$.
 	\item Prepare a single particle in a box, the volume of which is set to one.
 	\item Insert a partition, dividing the box into two parts of volume $v_y$ and $1-v_y$.
 	\item Slowly move the partition towards the middle of the box until the two parts are of equal volume.
 \end{enumerate}
 Importantly, there is no measurement and feedback in this procedure. The resulting probability distribution will be denoted $P^{\rm MI}_{\rm B}(x,y)$, where $y$ is determined probabilistically in step 1 and $x$ denotes the location of the particle after step 3. The work extracted in this process for given $x$ and $y$ is equal to $-W(x,y)$.
 
 The probability distributions of the forward and the backward experiment fulfill a fluctuation relation
\begin{equation}
\label{eq:frmi}
\frac{P_{\rm B}^{\rm MI}(x,y)}{P(x,y)}=e^{\beta W(x,y)-I(x,y)}.
\end{equation}
Here $I(x,y)$ is an information term
\begin{equation}
\label{eq:mi}
I(x,y) = \ln\frac{P(x,y)}{P(x)P(y)},
\end{equation}
such that if averaged over the forward distribution $P(x,y)$, it produces the mutual information between $x$ and $y$. Here we introduced the marginal distribution $P(x)=\sum_y P(x,y)$. We now define the continuous probability distributions
\begin{equation}
\label{eq:pwi}
P(W,I)=\sum_{x,y}\delta[W-W(x,y)]\delta[I-I(x,y)]P(x,y),
\end{equation}
and
\begin{equation}
\label{eq:pbwi}
P_{\rm B}^{\rm MI}(W,I)=\sum_{x,y}\delta[W+W(x,y)]\delta[I+I(x,y)]P^{\rm MI}_{\rm B}(x,y).
\end{equation}
We note that this equation implies that we assign the work value $-W(x,y)$ and the information value $-I(x,y)$ to a backward experiment described by the variables $x$ and $y$. While for the work, this corresponds to the actual extracted work, the information term is not directly related to the mutual information in the backward experiment.
With these definitions, the fluctuation relation in Eq.~\eqref{eq:frmi} implies
\begin{equation}
\label{eq:frmi2}
\frac{P^{\rm MI}_{\rm B}(-W,-I)}{P(W,I)}=e^{\beta W-I},
\end{equation}
which is of the form of Eq.~\eqref{eq:flucrel} with $\phi=W$ and $\sigma_{\rm I}=-\beta W+I$. A TUR of the form of Eq.~\eqref{eq:gturfb} thus holds, where the backward averages are taken over the distribution $P_{\rm B}^{MI}$, see Fig.~\ref{fig:szilard}.

The second backward experiment was introduced by the authors in Ref.~\cite{potts:2018}. It includes the same steps as the last backward experiment, as well as an additional post-selection step
\begin{enumerate}
	\setcounter{enumi}{4}
	\item Measure the location of the particle. If the outcome is equal to $y$ (determined in step 1), the experiment is considered successful. Otherwise, restart the experiment from step 2 (with the same value for $y$).
\end{enumerate}
This procedure is described by the probability distribution $P_{\rm B}^{\rm IE}(x,y)$. For this backward experiment, a fluctuation relation can be derived
\begin{equation}
\label{eq:frie}
\frac{P_{\rm B}^{\rm IE}(x,y)}{P(x,y)}=e^{\beta W(x,y)+\mathcal{E}(y)}.
\end{equation}
Denoting the success probability in step 5 by $P_{\rm S}(y)$, we write
\begin{equation}
\label{eq:infent}
\mathcal{E}(y)=\ln\frac{P(y)}{P_{\rm S}(y)},
\end{equation}
which can be interpreted as the entropy production that is inferable from the measurement outcome alone \cite{potts:2018}.
In complete analogy to Eq.~\eqref{eq:frmi2}, we find
\begin{equation}
\label{eq:frie2}
\frac{P^{\rm IE}_{\rm B}(-W,-\mathcal{E})}{P(W,\mathcal{E})}=e^{\beta W+\mathcal{E}},
\end{equation}
which is of the form of Eq.~\eqref{eq:flucrel} with $\phi=W$ and $\sigma_{\rm I}=-\beta W-\mathcal{E}$.

The TURs resulting from the two fluctuation relations are illustrated in Fig.~\ref{fig:szilard}. Depending on the parameters, either backward experiment can result in a tighter bound on the signal-to-noise ratio of the extracted work. In the limit of an error-free measurement (i.e, $\xi=0$ or $\xi=1$), the bound obtained from the inferable entropy is tight. This reflects the fact that the total entropy production can be inferred from the measurement. Explicit expressions for the probability distributions, the work information terms, as well as the averages involved in the TUR are given in App.~\ref{app:szilard}.

\section{Conclusions}
We showed how any fluctuation relation implies a TUR. The obtained relation does not constrain the forward experiment alone but includes both the forward as well as the backward experiment. This implies that the conventional TUR can be overcome in processes where a backward experiment compensates for the high signal-to-noise ratio in the forward experiment. Our results allow for directly extending the rich variety of fluctuation relations to TURs. In particular, we provide examples of TURs in the scenario of work extraction using measurement and feedback in a Szilard engine.

We note that independent related results have been obtained simultaneously. Reference \cite{vu:2019inf} considers measurement and feedback scenarios satisfying $P_{\rm B}=P$ such that the results of Ref.~\cite{hasegawa:2019} can be applied. Reference \cite{timpanaro:2019} obtains the tightest possible generalized TUR that can be obtained from a fluctuation relation with $P_{\rm B}=P$. Using a geometric approach, Ref.~\cite{falasco:2019} obtains novel results on TURs and recovers a number of recent results, including the ones from Refs.~\cite{hasegawa:2019,proesmans:2019}.

We thank Y. Hasegawa for valuable feedback. This work was supported by the Swedish Research Council. P.P.P. acknowledges funding from the European Union's Horizon 2020 research and innovation programme under the Marie Sk{\l}odowska-Curie Grant Agreement No. 796700.

\bibliography{biblio}

\appendix

\section{Quasistatic work extraction}
\label{app:qdot}
Consider a single realization of the quantum dot example discussed in the main text. Let $x_t=1$ ($x_t=0$) if the dot is occupied (empty) at time $t$. In the quasistatic limit, a time scale $\theta$ exists such that
\begin{equation}
\label{eq:qslimit}
\varepsilon_t\simeq\varepsilon_{t+\theta}\hspace{.25cm}\Rightarrow\hspace{.25cm} f(\varepsilon_t)\simeq f(\varepsilon_{t+\theta}),\hspace{.5cm}\Gamma\theta\gg 1,
\end{equation}
where $\Gamma$ denotes the coupling between the dot and the bath (i.e., in the absence of a drive, the dot thermalizes over the time-scale $1/\Gamma$). In this case, we find
\begin{equation}
\label{eq:insttherm}
\frac{1}{\theta}\int_{t}^{t+\theta}x_{t'}dt' \simeq f(\varepsilon_t), 
\end{equation}
where we use the fact that the time-average is equal to the ensemble average in the absence of a drive. The work extracted from the trajectory $x_t$ is then
\begin{equation}
\label{eq:workqs}
\begin{aligned}
&W=-\int_0^\tau dt \dot{\varepsilon}_t x_t\simeq-\int_0^\tau dt \dot{\varepsilon}_tf(\varepsilon)\\& = \int_{\varepsilon_\tau}^{\varepsilon_0} d\varepsilon f(\varepsilon)=-k_{B}T\ln\frac{1+e^{-\beta\varepsilon_0}}{1+e^{-\beta\varepsilon_\tau}}=\Delta F,
\end{aligned}
\end{equation}
where $\varepsilon_\tau=\varepsilon_0-\Delta \varepsilon$ and $\Delta F=F(\varepsilon_0)-F(\varepsilon_\tau)$ is the free energy difference between the thermal states at the initial and the final energies. Here we used
\begin{equation}
\label{eq:freeen}
F(\varepsilon) = -k_{B}T\ln Z(\varepsilon),\hspace{1cm}Z(\varepsilon)=1+e^{-\beta\varepsilon}.
\end{equation}
In the quasistatic limit, the maximal amount of work $\Delta F$ is thus extracted in each experimental run. The work distribution therefore tends to a Dirac delta distribution, see Eq.~\eqref{eq:qdotstat}. An analogous reasoning applies for the backward experiment.

\section{Szilard engine}
\label{app:szilard}
As discussed in the main text, we consider a particle in a box of volume $v=1$. Starting in thermal equilibrium, the particle is equally likely to be found in the left and in the right half of the box. A partition (wall) is then inserted in the middle of the box and a measurement of the position of the particle is performed. We denote the location of the particle by $x=\rm L,R$ and the measurement outcome by $y=\rm l,r$. We assume that a measurement error happens with probability $\xi$. The joint probability for $x$ and $y$ reads
\begin{equation}
\label{eq:probajointszilard}
P(x,y)=\delta_{x,y}(1-\xi)/2+(1-\delta_{x,{y}})\xi/2,
\end{equation}
where the Kronecker delta is defined as $\delta_{\rm L,l}=\delta_{\rm R,r}=1$ and zero otherwise. The marginals of this distribution read $P(x)=P(y)=1/2$. Having measured $y$, the partition is then moved away from where the particle is assumed to be, extending the volume it presumably occupies to $v_y\leq 1$. 
To evaluate the work extracted in this procedure, we consider the single particle as an ideal gas, described by $k_BT=pv$
where $p$ is the pressure and $v$ the volume. The extracted work is then given by
\begin{equation}
\label{eq:workgas}
W=\int pdv,
\end{equation}
resulting in
\begin{equation}
\label{eq:workszilard}
\begin{aligned}
\beta W(x,y)=\ln(2)&+\delta_{x,y}\ln(v_y)\\&+(1-\delta_{x,{y}})\ln(1- v_{y}).
\end{aligned}
\end{equation}
The first two moments of the work then read
\begin{equation}
\label{eq:mwsz}
\beta\langle W\rangle = \ln(2)+\frac{1}{2}\sum_{y=\rm l,r}\left[(1-\xi)\ln(v_y)+\xi\ln(1-v_y)\right],
\end{equation}
and
\begin{equation}
\label{eq:smwsz}
\beta^2\langle W^2\rangle = \frac{1}{2}\sum_{y=\rm l,r}\left[(1-\xi)\ln^2(2v_y)+\xi\ln^2(2-2v_y)\right].
\end{equation}
The variance of the work vanishes in the limit $\xi\rightarrow 0$ and $v_{\rm l}=v_{\rm r}$, since every run produces the same amount of work in this case. Similarly, the variance vanishes in the trivial case $v_y=1/2$, which corresponds to not doing anything after inserting the partition. We will now consider the two different backward experiments introduced in the main text.

\subsection{Mutual Information}
Here we consider the relation put forward in Refs.~\cite{sagawa:2010,sagawa:2012}. The probability distribution for this backward experiment (see main text for details) reads
\begin{equation}
\label{eq:pbacksusz}
P_{\rm B}(x,y)=\delta_{x,y} v_y/2+(1-\delta_{x,{y}})(1-v_y)/2,
\end{equation}
resulting in the fluctuation relation given in Eq.~\eqref{eq:frmi} with
\begin{equation}
\label{eq:miexp}
I(x,y) =\ln(2)+\delta_{x,y}\ln(1-\xi)+(1-\delta_{x,{y}})\ln(\xi).
\end{equation}
We further find the averages
\begin{equation}
\label{eq:mwbsusz}
\begin{aligned}
\beta\langle W\rangle_{\rm B} &= -\sum_{x,y}W(x,y)P_{\rm{B}}(x,y)=-\ln(2)\\&-\frac{1}{2}\sum_{y=\rm l,r}\left[v_y\ln(v_y)+(1-v_y)\ln(1-v_y)\right],
\end{aligned}
\end{equation}
and
\begin{equation}
\label{eq:smwbsusz}
\beta^2\langle W^2\rangle_{\rm B} = \frac{1}{2}\sum_{y=\rm l,r}\left[v_y\ln^2(2v_y)+(1-v_y)\ln^2(2-2v_y)\right].
\end{equation}
For the information terms, we find
\begin{equation}
\label{eq:infsusz}
\langle I\rangle = \ln(2)+\frac{1}{2}\sum_{y=\rm l,r}\left[(1-\xi)\ln\left(1-\xi\right)+\xi\ln\left(\xi\right)\right],
\end{equation}
which is the mutual information in the forward experiment and
\begin{equation}
\label{eq:infbsusz}
\langle I\rangle_{\rm B} = -\ln(2)-\frac{1}{2}\sum_{y=\rm l,r}\left[v_y\ln\left(1-\xi\right)+(1-v_y)\ln\left(\xi\right)\right],
\end{equation}
which is \textit{not} the mutual information in the backward experiment. Note that in the backward experiment, the work is $-W(x,y)$ and the information $-I(x,y)$.

\subsection{Inferable entropy production}
Here we consider the fluctuation relation put forward in Ref.~\cite{potts:2018}. The probability distribution for this backward experiment (see main text for details) reads
\begin{equation}
\label{eq:pbackszilard}
P_{\rm B}(x,y)=\frac{1}{2}\frac{\delta_{x,y}v_y(1-\xi)+(1-\delta_{x,{y}})\xi(1-v_y)}{v_y(1-\xi)+\xi(1-v_y)},
\end{equation}
resulting in the fluctuation relation given in Eq.~\eqref{eq:frie} with
\begin{equation}
\label{eq:sigmaiinfsz}
\mathcal{E}(y) =-\ln[2v_y(1-\xi)+2\xi(1-v_y)],
\end{equation}
where the success probability introduced in the main text reads $P_{\rm S}(y) = v_y(1-\xi)+\xi(1-v_y)$.
We further find the averages
\begin{equation}
\label{eq:mwbsz}
\begin{aligned}
&\beta\langle W\rangle_{\rm B} = -\ln(2)\\&-\frac{1}{2}\sum_{y=\rm l,r}\frac{(1-\xi)v_y\ln(v_y)+\xi(1-v_y)\ln(1-v_y)}{(1-\xi)v_y+\xi(1-v_y)},
\end{aligned}
\end{equation}
and
\begin{equation}
\label{eq:smwbsz}
\begin{aligned}
&\beta^2\langle W^2\rangle_{\rm B} = \frac{1}{2}\times\\&\sum_{y=\rm l,r}\frac{(1-\xi)v_y\ln^2(2v_y)+\xi(1-v_y)\ln^2(2-2v_y)}{(1-\xi)v_y+\xi(1-v_y)}.
\end{aligned}
\end{equation}
For the inferable entropy production, we find
\begin{equation}
\label{eq:infsz}
\langle \mathcal{E}\rangle =-\langle \mathcal{E}\rangle_{\rm B} = -\frac{1}{2}\sum_{y=\rm l,r}\ln[2v_y(1-\xi)+2\xi(1-v_y)].
\end{equation}
Note that the first equality sign holds in this example but is not generally true.

\end{document}